\documentclass{article}
\usepackage{spconf}
\usepackage{amsmath}
\usepackage{newtxtext,newtxmath}
\usepackage{graphicx}
\usepackage{booktabs}
\usepackage{cite}
\usepackage[hidelinks]{hyperref}

\graphicspath{{figures/}}

\begin{document}
\ninept

\title{MuseTimbre: Zero-Shot Timbre Transfer by Controlling a Frozen Music Generator}

\twoauthors
  {Yuan-Chiao Cheng}
  {Music Informatics Group \\ Georgia Institute of Technology \\ Atlanta, GA, USA}
  {Zhiyao Duan}
  {Audio Information Research (AIR) Lab\\ University of Rochester \\ Rochester, NY, USA}

\maketitle

\begin{abstract}
Instrument timbre transfer re-voices a performance using the timbre of another instrument. Extracting the target timbre from an audio reference capture more nuances than inferring it from a text prompt. Systems that read timbre from such a clip train a dedicated model for the task, which captures the timbre cleanly but stays a narrow, single-purpose system. More versatile approaches add control to a pretrained music generator, yet a reference clip entangles timbre with genre and melody, so these systems fall back on text to name the timbre.
We present MuseTimbre, the first system, to our best knowledge, that transfers timbre from an audio reference to a polyphonic source through conditioning a pretrained music generator.
This system employs a multi-pitch estimator to extract pitch information from the source and finetune a CLAP encoder to extract timbre information from the reference audio.
Experiments show that across four datasets of real polyphonic recordings, MuseTimbre achieves pitch alignment on par with the baselines while matching the reference timbre far more closely. Results also show that the finetuned CLAP-based timbre extractor is robust to pitch variations, making it useful in timbre similarity measures.
\end{abstract}


\section{Introduction}

Musicians and producers often want to re-voice an existing performance with a chosen instrument while keeping its notes. This task, \emph{instrument timbre transfer}, converts a source recording to the timbre of a reference instrument while preserving the musical content. The reference timbre could be described by a text prompt containing the instrument name, but this approach is typically not specific enough on timbral nuances~\cite{musecontrollite25,adatt26}. This is because different instruments of the same type can cover a large timbral variety, e.g., clean jazz tone vs. heavy distortion for the electric guitar, or bright soloist vs. warm orchestral tone for the violin. The most precise way to provide the reference timbre, undoubtedly, is to provide an audio reference~\cite{cifka21ssvqvae,demerle24combining,kim2025tokensynth}.
Integrating the timbre extracted from the audio reference with the pitch information extracted from the source audio requires good disentanglement and fusion of timbre and pitch information.

\begin{figure*}[t]
\centering
\includegraphics[width=\textwidth]{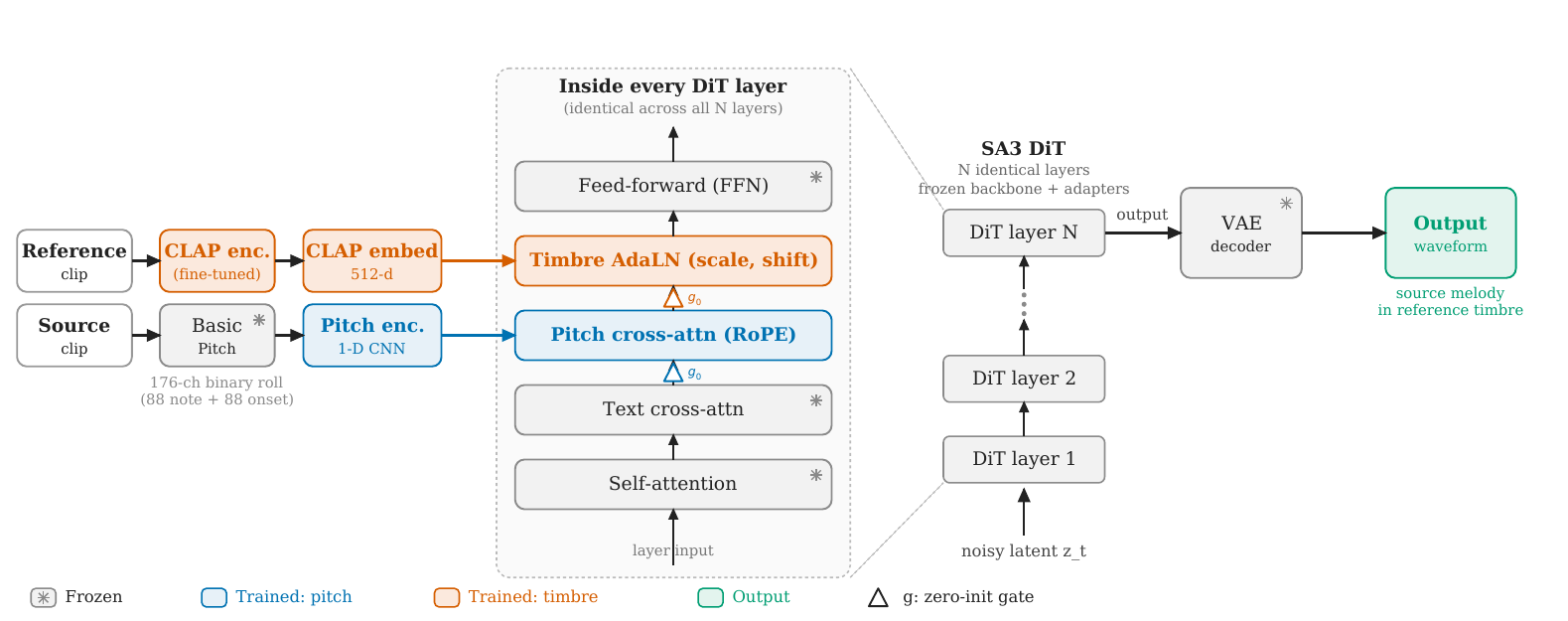}
\caption{System overview. We freeze Stable Audio 3 Medium, its autoencoder, and its text control (gray) and add two trained modules, pitch (blue) and timbre (orange). A source clip or MIDI is encoded into a binary note-and-onset roll; a reference clip is encoded into a time-invariant embedding from a fine-tuned CLAP encoder. Pitch and timbre enter through cross-attention and AdaLN, respectively.}
\label{fig:system}
\end{figure*}

The first challenge is timbre-pitch disentanglement from the source recording.
Prior work handles it with varied explicitness.
Diffusion inversion methods gradually remove timbral information from the source, but may also corrupt its musical content because the two are not explicitly separated \cite{lee2026mitimbre,baoueb2025difftone,manor24zeta}.
Self-supervised methods instead compress the source through a discrete bottleneck and rely on it to discard attributes beyond musical content \cite{cifka21ssvqvae}.
Other methods attach supervised controls for musical attributes to a pretrained generator, leaving timbre to be modified through text \cite{musecontrollite25,wu24musiccontrolnet,floresgarcia25sketch2sound,adatt26}.
The most explicit approach trains a dedicated model to separate pitch from timbre and reconstruct audio from the two conditions \cite{engel20ddsp,demerle24combining,kim2025tokensynth}.

The second challenge is to specify and transfer the target timbre precisely.
Text-conditioned generators use the target instrument name~\cite{musecontrollite25,adatt26}, which cannot specify its finer timbral nuances.
Audio-reference methods capture such nuances more directly, but train dedicated generators for a narrow or fixed instrument set~\cite{cifka21ssvqvae,demerle24combining,kim2025tokensynth}.

In this paper, we propose MuseTimbre, a system for instrument timbre transfer from an audio reference. Using two plug-in control modules on a pretrained music generator, Stable Audio 3 Medium \cite{stableaudio3_26}, in the spirit of decoupled adapters for image generation \cite{ipadapter23}, we address both challenges.
For the first challenge, we use Basic Pitch \cite{bittner22basicpitch} to estimate the pitch information of the source explicitly as a piano roll, which discards the timbre information of the source signal. The pitch information is then encoded by a 1-D CNN and fed to the music generator through cross-attention.
For the second challenge, we use an audio-text encoder, LAION-CLAP \cite{laionclap23}, to extract a timbre embedding from the audio reference and inject it into the pretrained music generator through adaptive layer normalization.
Unlike prior reference-conditioned systems that keep this encoder frozen \cite{kim2025tokensynth,ipadapter23}, we fine-tune it end to end on the transfer task.
Through this fine-tuning, the encoder also learns to recognize an instrument across changing pitch and content, which we verify with a timbre retrieval task.

The system overview is shown in Figure~\ref{fig:system}. All trainable modules are trained together on our task.
This design brings three properties. The multi-pitch estimator lets the system process polyphonic sources. The timbre encoder and the music generator are both pretrained on large-scale diverse audio, so the system transfers the timbre of a large variety of instruments. The piano-roll input also accepts a MIDI score directly, so the system can synthesize a score in the timbre of a reference clip, with the generator's original text control applied on top. We evaluate on real polyphonic recordings from four datasets that are unseen during training. Results show that MuseTimbre matches the reference timbre far more closely than every baseline while staying competitive on pitch preservation.

In contrast to prior work, our contributions are twofold. First, MuseTimbre is the first system that transfers timbre from an audio reference to a polyphonic source through conditioning a pretrained music generator. Two separate conditioning modules encode the pitch from the source and timbre from the reference respectively.
Second, fine-tuning the CLAP encoder on this task turns it into a pitch-robust timbre encoder, confirmed by a timbre retrieval test, which can serve as a timbre-similarity metric for future evaluation works. We release our code, model weights, and audio examples.\footnote{Code: \url{https://github.com/barry-mir/muse-timbre}; audio: \url{https://barry-mir.github.io/muse-timbre-demo/}}

\section{Method}

\subsection{Scope and definitions}
\label{sec:scope}
\emph{Instrument timbre transfer} is the task of simulating what happens when the \emph{music content} of a source recording is played by the \emph{timbre} of a different instrument.
When restricting to harmonic instruments, the music content can be characterized by the temporal organization of \emph{pitch}. Therefore, in this paper, pitch and content refer to the same concept.
For polyphonic music, there can be multiple pitches at a time, and a common way to represent it is a piano roll representation.

Timbre is the acoustic identity of notes; it is related to the spectral envelope and other characteristics like note articulation, and it can be used to distinguish different instruments.
From the source-filter perspective of sound synthesis~\cite{engel20ddsp}, pitch characterizes the source while timbre characterizes the resonance filter of an instrument.

Pitch and timbre are related by nature; an instrument's spectral envelope may change with the pitch register, and human pitch perception depends on timbre. Therefore, learning disentangled representations of pitch and timbre from a polyphonic music recording is challenging. In this paper, instead, we employ a pitch estimator and a CLAP encoder to explicitly extract pitch and timbre from the source and reference, respectively.

\subsection{System overview}
Figure~\ref{fig:system} illustrates MuseTimbre. Rather than training a generative model from scratch, we add two conditioning pathways to Stable Audio 3 Medium~\cite{stableaudio3_26}, a pretrained rectified-flow diffusion transformer (DiT) for music generation. Its audio autoencoder, DiT backbone, and original text-conditioning pathway remain frozen. The two new pathways provide different forms of control to the frozen generator. The \emph{pitch conditioning} pathway converts the source into a binary piano roll representation and injects it into the DiT through cross-attention. The \emph{timbre conditioning} pathway converts the reference audio into a single time-invariant embedding and uses it to modulate the DiT features through adaptive layer normalization. The following subsections describe the two pathways in detail.

\subsection{Pitch conditioning}
\label{sec:pitch}
The pitch condition is a binary piano roll with 88 note and 88 onset channels, resampled to the latent frame rate of about 10.8 frames per second. The note channel denotes the frames a note lasts while the onset channel denotes the first frame of a note, and silence is zero in both.
We use Basic Pitch \cite{bittner22basicpitch} to transcribe the piano roll of source audio. We binarize the continuous transcription salience at a fixed threshold, 0.35 for notes and 0.50 for onsets, tuned to balance precision and recall. This removes the harmonic structure, which correlates with the source timbre, so the pitch channel carries note content with almost no timbre information. Newer transcribers such as MuScriptor are more accurate \cite{muscriptor2026}, but we reserve MuScriptor for evaluation and keep Basic Pitch inside the model for fairness.
A 1-D CNN $g_p$ encodes the roll into per-frame features. They enter every layer of the DiT backbone through decoupled cross-attention with rotary embeddings on queries, keys, and values, behind a zero-initialized gate.

\begin{figure*}[t]
\centering
\begin{minipage}[t]{0.48\textwidth}
\centering
\includegraphics[width=\linewidth]{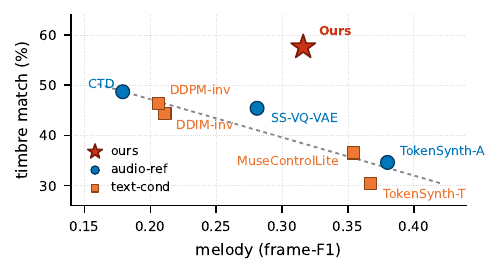}
\caption{Zero-shot transfer on 304 pairs: melody (frame-F1) against timbre match (PaSST); upper-right is better. TokenSynth takes a Basic Pitch transcription.}
\label{fig:scatter}
\end{minipage}\hfill
\begin{minipage}[t]{0.48\textwidth}
\centering
\includegraphics[width=\linewidth]{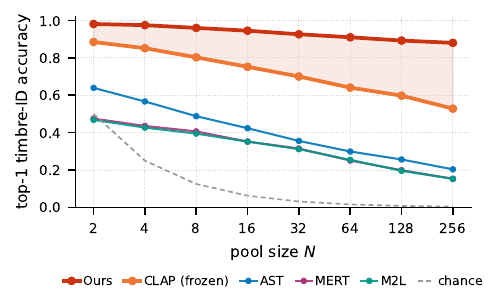}
\caption{Timbre-identification retrieval on the controlled grid. Fine-tuned CLAP (Ours) tops frozen CLAP and the other baselines at every pool size; chance is $1/N$.}
\label{fig:retrieval}
\end{minipage}
\end{figure*}

\subsection{Timbre conditioning}
\label{sec:timbre}
The timbre encoder $f_t$ is the audio branch of LAION-CLAP trained on music data \cite{laionclap23}. It reads the raw reference waveform sampled at 48 kHz and returns one time-invariant 512-dimensional embedding.
Because the timbre embedding is a single global control, we inject it into the DiT through adaptive layer normalization (AdaLN) instead of cross-attention. A linear projection maps the embedding $x$ to a per-block scale $\gamma$ and shift $\beta$, formulated as $x \leftarrow x\,(1+\gamma) + \beta$. This adaptive layer norm sits after the pitch cross-attention in each layer, behind a zero-initialized gate in the AdaLN-Zero style \cite{peebles2023dit}.

\subsection{Training objective and disentanglement}
We train the whole system with the rectified-flow objective. Given a target latent $z_0$ and Gaussian noise $\epsilon$, we form $x_t = (1-t)\,z_0 + t\,\epsilon$ and regress the velocity toward $\epsilon - z_0$,
\begin{equation}
\mathcal{L} = \mathbb{E}_{z_0,\epsilon,t}\,\big\lVert\, v_\theta(x_t, t, c_{\text{pitch}}, c_{\text{timbre}}) - (\epsilon - z_0) \,\big\rVert_2^2 ,
\end{equation}
where $v_\theta(\cdot)$ is the predicted velocity at diffusion time step $t$, and $c_{\text{pitch}}$ and $c_{\text{timbre}}$ denote the pitch and timbre conditions, respectively.
For each training clip, $c_{\text{pitch}}$ is either the ground-truth pitch or the estimated pitch by Basic Pitch followed by encoding steps described in Section~\ref{sec:pitch}, while $c_{\text{timbre}}$ is extracted from a different segment of the same clip using the timbre extraction procedure described in Section~\ref{sec:timbre}.
We train on a 50/50 mix of synthesized audio and real audio.
For the synthesized half, we render Slakh MIDI stems~\cite{manilow2019slakh} with seven General MIDI soundfonts, yielding approximately 1.1 million stem recordings of at least ten seconds across thirteen instrument classes. The real half contains 18,869 clips from 1,208 single-instrument stems of MoisesDB and URMP \cite{pereira23moisesdb,li18urmp}, with drums, percussion, vocals, and untagged stems dropped, and supplies binarized Basic Pitch rolls. Every training clip is five seconds at 44.1 kHz. We optimize with AdamW for 250k steps on two GPUs, with weight decay 0.01, 1000 warmup steps then cosine decay, effective batch size 16, and multi-condition classifier-free guidance (CFG) dropout.
The learning rate for the timbre encoder is $10^{-5}$ as it was already pretrained, while the learning rate for the control modules is $10^{-4}$.

\subsection{Inference}
Inference starts from noise and runs 25 Euler steps, taking about one second on a single RTX 5090 GPU for five seconds of audio. We combine the conditions with CFG $\lambda_{\text{pitch}} = \lambda_{\text{timbre}} = 2$, which keeps both pitch and timbre match high. The frozen backbone keeps two further capabilities: 1) The system can take a score piano roll as the input to the pitch conditioning module and render it with the timbre of the reference clip; 2) The pretrained text control of the synthesis model is untouched and still works alongside the pitch and timbre conditioning. On the demo page we demonstrate its use for the acoustic environment, such as room and reverb.

\section{Experiments}

\begin{figure*}[t]
\centering
\begin{minipage}[c]{\columnwidth}
\centering
\makeatletter\def\@captype{table}\makeatother
\caption{Pitch-timbre separation by module ablation on the 304 pairs at $\lambda_{pitch} = \lambda_{timbre} = 2$. Removing the timbre module collapses timbre while melody largely holds, and removing the pitch module collapses melody while timbre stays, a double dissociation.}
\label{tab:ablation}
\setlength{\tabcolsep}{6pt}
\begin{tabular}{@{}lcc@{}}
\toprule
Conditioning & Timbre match (\%)\,$\uparrow$ & frame-F1\,$\uparrow$ \\
\midrule
Pitch + timbre (full) & 57.6 & 0.316 \\
Pitch only (null timbre) & 16.1 & 0.267 \\
Timbre only (null pitch) & 74.7 & 0.041 \\
\bottomrule
\end{tabular}
\end{minipage}\hfill
\begin{minipage}[c]{\columnwidth}
\centering
\includegraphics[width=0.62\linewidth]{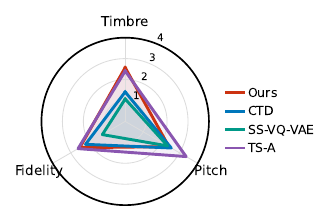}
\caption{Listening test, mean rating from 0 to 5 per system on timbre, pitch, and fidelity over twenty listeners and ten clips.}
\label{fig:listening}
\end{minipage}
\end{figure*}

\begin{table}[t]
\centering
\scriptsize
\setlength{\tabcolsep}{3pt}
\renewcommand{\arraystretch}{0.85}
\caption{AudioSet labels counted as a timbre match.}
\label{tab:families}
\begin{tabular}{@{}ll@{}}
\toprule
Reference & Labels counted as a hit \\
\midrule
violin, viola & Violin; Bowed string; String section \\
cello, double bass & Cello / Double bass; Bowed string; String section \\
flute, clarinet, saxophone & own label; Woodwind \\
oboe, bassoon & Woodwind \\
trumpet, horn & Trumpet / French horn; Brass \\
piano & Piano; Keyboard \\
electric guitar & Electric guitar; Guitar; Plucked string \\
\bottomrule
\end{tabular}
\end{table}

\subsection{Setup}
\label{sec:setup}
We evaluate zero-shot cross-instrument transfer on 304 pairs from four datasets with ground-truth MIDI.
We hold all four out of training: PHENICX-Anechoic \cite{miron2016phenicx}, the MAESTRO v3 test split \cite{hawthorne2019maestro}, GOAT \cite{loth2025goat}, and Bach10 \cite{duan2010bach10}, thirteen instruments in total. We draw isolated single-instrument clips and take the loudest five seconds of each to avoid silence. We pair each source with a reference of a different instrument from a different dataset, so no pair shares instrument identity or recording condition.

We compare against four text-conditioned and three audio-reference baselines. \textbf{DDIM-inv} \cite{song2021ddim} and \textbf{DDPM-inv} invert the source under a source prompt and re-denoise under a target prompt, the latter following ZETA \cite{manor24zeta}, and both share our Stable Audio 3 backbone. \textbf{MuseControlLite} conditions on a source melody signal and a target-instrument prompt on Stable Audio Open \cite{musecontrollite25}. \textbf{TokenSynth-T} is TokenSynth conditioned on a CLAP text embedding of the target instrument \cite{kim2025tokensynth}. In the audio-reference group, \textbf{Control-Transfer Diffusion (CTD)} is the nearest method to ours: it also conditions a diffusion model on a pitch signal and an audio timbre embedding, but trains a standalone model instead of controlling a pretrained one \cite{demerle24combining}. We run its official real-instruments checkpoint. \textbf{TokenSynth-A} is TokenSynth conditioned on a CLAP audio embedding of the reference \cite{kim2025tokensynth}. \textbf{SS-VQ-VAE} is a spectrogram autoencoder that splits each clip into pitch and timbre embeddings, then decodes the source's pitch embedding with the reference's timbre embedding at 16 kHz \cite{cifka21ssvqvae}. Both TokenSynth variants take the same Basic Pitch transcription of the source that our model uses, ensuring that differences do not arise from the pitch input.

We report two metrics. \emph{Timbre match} uses PaSST, an audio tagger trained on AudioSet \cite{koutini22passt}: we count a prediction as a hit when the top-1 class of the output falls within the reference instrument's family. Each of the thirteen instruments maps to its AudioSet label and its parents (Table~\ref{tab:families}), so all methods are evaluated consistently at the family level. \emph{Pitch preservation} is measured by the frame-F1 between a transcription of the output and the ground-truth source MIDI, at a 10 ms hop and 50 cent tolerance, over all 304 pairs. We transcribe with MuScriptor conditioned on the target instrument \cite{muscriptor2026}. It is noted that both metrics have an upper bound. In particular, reference clips achieve a PaSST accuracy of 88.5\%, and transcribing the source audio achieves an average frame-F1 of 0.540.

\subsection{Main results}
Figure~\ref{fig:scatter} plots every method by pitch match against timbre match. The baselines exhibit a trade-off between pitch preservation and timbre match: methods with higher frame-F1 generally achieve lower timbre match. We match the reference timbre 57.6 percent of the time at frame-F1 0.316. The audio-reference baselines trade one axis for the other: CTD and SS-VQ-VAE reach 48.7\% and 45.4\% on timbre match, but lower frame-F1. The two TokenSynth variants reach higher frame-F1, 0.380 and 0.367, but lower timbre match, and within TokenSynth the audio reference beats the text prompt, 34.6\% for TokenSynth-A against 30.4\% for TokenSynth-T, showing the benefit of using audio as timbre reference as we assumed earlier.

\subsection{A pitch-robust timbre encoder}
Training the transfer task end to end reshapes the CLAP embedding into a timbre-similarity space. To measure this we build a controlled grid: thirteen instruments crossed with forty melodic segments from the four datasets, and we render every combination of timbre and content. Each retrieval pool holds one target, the query's instrument with different note content, and $N{-}1$ distractors, for pool sizes $N$ from 2 to 256. At least one distractor plays the query's exact notes in a different timbre. A content encoder that ignores the timbre ranks that distractor first, and a pitch-robust timbre encoder ranks the target first (Figure~\ref{fig:retrieval}). Fine-tuning lifts the frozen CLAP encoder from 0.53 to 0.88 top-1 accuracy at the largest pool, a 35-point gain that widens as the pool grows. Both sit far above three pretrained audio embeddings, vanilla AST \cite{gong21ast}, MERT \cite{li2023mert}, and Music2Latent \cite{pasini24music2latent}. High accuracy here shows our fine-tuned encoder is a candidate pitch-robust timbre-similarity metric. We test on synthetic renders because they are controllable, since real recordings rarely offer the same content in different timbres, and we leave further timbre-similarity evaluation to future work.

\subsection{Subjective evaluation}
We run a listening test on the audio-reference systems: Ours, CTD, SS-VQ-VAE, and the Basic-Pitch-fed TokenSynth-A. The design follows MUSHRA (multiple stimuli with hidden reference) without anchors \cite{itu2015mushra}. Twenty listeners with musical training hear a content source and a timbre reference and then score four blinded outputs on timbre, pitch, and fidelity from 0 to 5, over ten tuples of source, reference, and systems (Figure~\ref{fig:listening}). Ours scores highest on timbre at 2.6, above TokenSynth-A at 2.4 and well above CTD at 1.4 and SS-VQ-VAE at 1.1. On fidelity ours stays within 0.1 of TokenSynth-A, 2.5 against 2.6. TokenSynth-A leads on pitch at 3.4, consistent with its higher melody score in Figure~\ref{fig:scatter}. This is consistent with the objective evaluation, where TokenSynth-A also achieves higher frame-F1.

\subsection{Pitch-timbre separation}
To test whether the two modules carry distinct information, we drop each condition at inference and re-evaluate the same pairs (Table~\ref{tab:ablation}). Unlike the previous evaluations, we use the same audio clip for timbre and pitch conditions and pass null embedding for the dropped condition. Dropping the timbre module collapses timbre match to 16.1\%, which is near the 11.8\% random guess rate. This is reasonable as Basic Pitch only extracts pitch but not timbre and the pitch range information only helps guess the instrument category slightly.
Dropping the pitch module does the opposite: timbre match maintains high and frame-F1 drops to 0.041. This suggests that the pitch information is not captured by the CLAP encoder or not utilized in the DiT decoding process.

\section{Conclusion}
We presented MuseTimbre, a zero-shot instrument timbre transfer system that conditions a pretrained music generator on the pitch content of a source recording and the timbre of an audio reference.
By representing pitch as a piano roll and reference timbre with a fine-tuned audio encoder, MuseTimbre adds explicit control over both attributes while keeping the pretrained generator frozen.
Across four held-out datasets, the system achieves substantially higher timbre match than the baselines while maintaining competitive pitch preservation.
Beyond generation, fine-tuning CLAP on the transfer task produces an embedding that identifies instrument timbre across changes in pitch content, suggesting its potential as a representation for timbre similarity.
Together, these results show that audio-reference timbre control can be added to a pretrained music generator without training a dedicated generator for the transfer task.


\end{document}